\documentclass[runningheads]{llncs}
\usepackage[T1]{fontenc}
\usepackage{graphicx}
\usepackage{amsmath}
\usepackage{graphicx}
\usepackage{adjustbox}
\usepackage{float}
\usepackage[title]{appendix}
\usepackage[colorlinks=true, allcolors=blue]{hyperref}
\usepackage{listings}
\usepackage{multicol}
\usepackage{color}
\usepackage{xcolor}
\usepackage{censor}

\begin{document}
\title{Accessible Design in Integrated Development Environments: A Think Aloud Study Exploring the Experiences of Students with ADHD}
\titlerunning{Accessible Design in Integrated Development Environments}
%
\author{Luke Halpin\inst{1,2}\orcidID{0009-0001-8878-2996} \and
Phillip Benachour\inst{1}\orcidID{0000-0001-8578-4024} \and 
Tracy Hall\inst{1}\orcidID{0000-0002-2728-9014} \and
Ann-Marie Houghton\inst{1}\orcidID{0000-0003-1289-1878} \and
Emily Winter\inst{1}\orcidID{0000-0003-3314-7300}
 }
\authorrunning{L. Halpin et al.}
%

\institute{Lancaster University, Lancaster, United Kingdom\\
\email{\{l.halpin,p.benachour,tracy.hall,a.houghton,e.winter\}@lancaster.ac.uk}
\and
Corresponding Author: Luke Halpin, l.halpin@lancaster.ac.uk} 

\maketitle              
\begin{abstract}
Coding forms a key part of computer science education in universities. As part of this education, Integrated Development Environments (IDEs) are essential tools for coding. However, it is currently unknown how the design of an IDE’s interface impacts on students with Attention Deficit Hyperactivity Disorder (ADHD). 

In this study we investigated the use of IDEs by students with ADHD. We conducted a think aloud study with nine university computing students, followed by qualitative observational interviews to analyse their learning and engagement with the Visual Studio Code IDE.
The paper reports on these experiences and seeks to understand the role IDEs play in the educational setting. 
 
Our work also examines how digital accessibility and usability are considered in the current design of IDEs. We analysed the qualitative data using a thematic analysis and identified three primary themes: self-confidence, interaction, and learning as well as various sub-themes. 
 
The themes and their sub-themes illustrate key areas of consideration when designing IDEs for students with ADHD. The primary findings highlight experiences of frustration and barriers in the current design and layout of IDEs. 
 
Through our participatory approach we provide a rare insight into ADHD user experiences around usability and accessibility, and describe the need for better design of development environments to ensure a positive learning experience for the students.

\end{abstract}
\section{Introduction}
Integrated Development Environments (IDEs) are a valuable tool for software developers, allowing them to use various integrated tools to improve the process of development. Yet IDE's design and poor usability, as demonstrated by Dillon \& Thompson\cite{dillon16}, can often be disconcerting to those at the early stages of learning the various skills associated with development and computer science \cite{valez2020}.

Investigating the experience of students with Attention Deficit Hyperactivity Disorder (ADHD) in computing education is especially important considering Kirdani-Ryan \& Ko's discussions around neurodivergent students being more likely to study computing \cite{kirdani24}. In addition to this fact, Leow et al. \cite{leow2024prevalence} found ADHD to have a strong correlation with students discontinuing their studies, which is already more prevalent in computing than other subjects \cite{hesanoncon}. ADHD is a developmental condition which impacts the regulation of executive functions and is often first noticed in childhood but is usually a lifelong condition \cite{BarkleyRussellA2018Ahd}.

In this exploratory study we invited undergraduate computer science students with ADHD to take part in a think aloud activity. Charters \cite{charters2003use} describes think aloud as a research method involving a participant vocalising their thought process while completing an activity to give insight into their interaction experience. The study activity in this paper consisted of students using an IDE to complete coding tasks then discussing their experiences using the IDE. This approach will address a current gap in the literature around the impact of IDE design on students with ADHD.

Our study was exploratory in nature, with the goal of developing an understanding of the experience of students with ADHD when using an IDE to complete coding tasks. The following research questions were formed to accomplish this goal:

\textbf{RQ1.} How does the design of the IDE user interface (UI) impact the usability for a student with ADHD?

\textbf{RQ2.} What are the experiences of students with ADHD when using an IDE?

RQ1 will allow aspects of current design to be identified that have an impact on students, while RQ2 will give insight into how students experience the overall interaction. These questions will be answered through a thematic analysis of the data collected during the think aloud activity and short interviews.

The contributions of this paper provide important insight on key challenges to making IDEs more accessible to students with ADHD and supporting them in their learning such as:
\begin{itemize}
    \item Supporting issues with self-confidence
    \item Managing a sense of overwhelm
    \item Improving communication between the IDE and the learner
\end{itemize}

The structure of this paper is as follows: Section 2 covers the background and related work giving an overview of current research around IDE usability and ADHD digital accessibility. Section 3 lays out the overall methodology including the design of the study and analysis of data. Section 4 breaks down the results of the analysis, explaining and evidencing the findings. Section 5 presents a discussion of the results and their implications while also addressing limitations. Section 6 contains the conclusion of our paper.

\section{Background and Related Work}

\subsection{IDEs and Usability}

While IDEs are a frequently utilised tool in software development \cite{latoza06}, the current research around their usability is limited. The work of Kline et al. \cite{kline2005evaluation} evaluates the usability of IDEs through interviews and heuristics evaluation and discusses how these tools are often centred on being designed for functionality rather than usability by developers. This design mentality can also lead to educational environments instead using slimmed down IDE variants such as BlueJ \cite{kolling2003bluej}. While slimmed down variants may be beneficial to learning environments through being simplified and easier to use, this can lead to skill gaps between what is expected at university level and what is expected in industry itself \cite{birillo24}.

Few of these studies on IDE usability focus on accessibility, with only a limited amount of research on ADHD and IDEs. Kasatskii et al. \cite{kasatskii2023effect}, for example, explored the role of perceptual load in the efficiency of software developers. They used a plugin to create different levels of perceptual load in the IDE to compare the performance of developers working in the differing environments, finding a positive impact of adaptability and need for individualised support and design. 

This paper however differs from existing work like that of Kasatskii et al. by examining the overall experience of using IDEs rather than just perceptual load and by examining an educational context rather than one in industry. 

\subsection{ADHD and Digital Accessibility Research}
 Spiel et al. \cite{spiel22} discuss the current body of work through a critical systematic literature review of ADHD and technology research. 
 The review highlighted a lack of research with ADHD users as primary participants, and a significant focus on tech to mitigate ADHD symptoms rather than support ADHD in context.
 The existing body of research focuses heavily on the current diagnostic criteria for ADHD, which has been criticised by researchers \cite{BarkleyRussellA2018Ahd}, and often implements a medical model approach. The medical model of disability is one which focuses on the disability being the condition and having a need to reduce the impact of the condition \cite{haegele2016disability}. An alternative and less explored model in ADHD research is the social model of disability where the barriers arising from an inaccessible environment are the problem that leads to the person with ADHD being disabled \cite{haegele2016disability}. The work of this paper addresses the gap in research by providing a social model perspective and considering the environmental barriers created by the IDE, in line with calls for such work by the likes of Spiel et al. \cite{spiel22}. By taking this approach participants give a full and detailed account of their holistic experience and how interfaces disable them, rather than focusing on specific aspects of their disabilities. This will identify the features of design which detriment the usability of the IDEs, and hence the obstacles that currently impact someone with ADHD. 
\section{Methodology}
\subsection{Study Design}
A retrospective think aloud protocol was followed. As part of the protocol, participants completed a coding task consisting of debugging and contributing original code to a provided code base. A retrospective protocol was chosen as it has been previously reported to provide more explanations and design information compared to a concurrent approach \cite{bowers1990concurrent}. A retrospective protocol also allows participants to have more focus on the coding task and the think aloud activity individually.

\begin{figure}[h]
    \centering
    \includegraphics[width=1\linewidth]{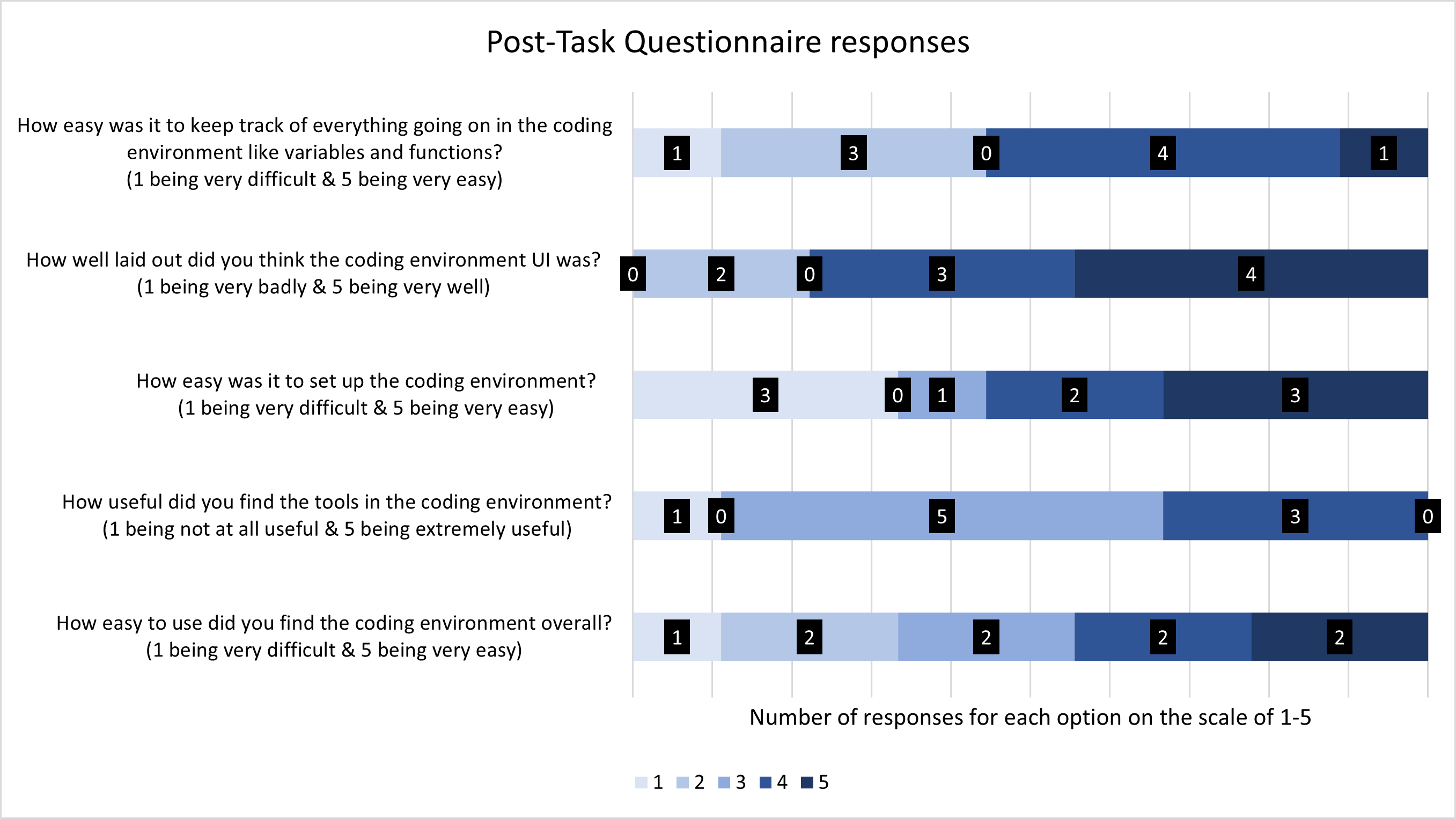}
    \caption{Participant responses to post-task questionnaire using Likert scales with 1 being a negative response and 5 being a positive response.}
    \label{likertgraph}
  \end{figure}

When designing the study, data triangulation was kept in mind to address some of the methods limitations \cite{charters2003use}. 
Firstly, by using the retrospective method, a lot of information that is sometimes lost in the concurrent variant is included with a lower cognitive load and gives more time for the participant to consider their thoughts. 
Secondly, the participants filled out a short questionnaire rating the five questions in Figure \ref{likertgraph} on Likert scales to provide some quantitative values and understand the general feeling of participants about the activity. 
Finally, the use of short informal interviews, which allowed us to discuss both the activity and questionnaire with the participants, and some of their wider experiences.

\subsection{Participants}
The participants recruited were undergraduates studying computer science related subjects. Participants ranged in age from 19 to 23 years old (19:1, 20:3, 21:3, 22:1, 23:1) with a majority female participants (male:3, female:6) and varying ethnic backgrounds (white:5, Black/African/Caribbean:1, Asian:3).

\begin{figure}[h]
    \centering
    \includegraphics[width=0.8\linewidth]{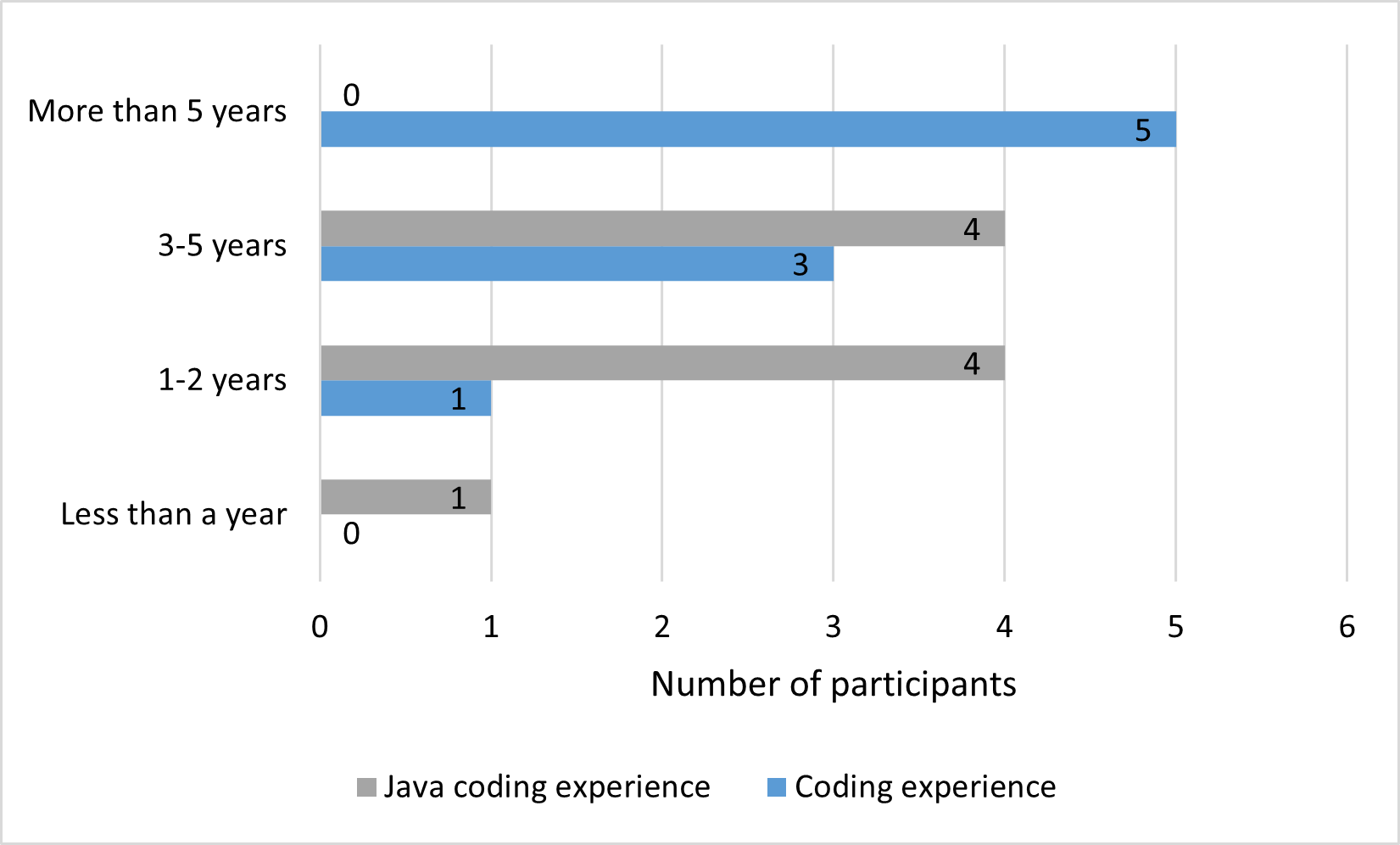}
    \caption{A bar chart showing the coding experience of participants.}
    \label{exptab}
\end{figure}

Participants were asked which IDEs they use most frequently, with the vast majority using VS Code (VS Code:7, Visual Studio:1, Jetbrains:1) and their experience coding was documented as shown in Figure \ref{exptab}.

For the study we required ADHD as an eligibility requirement. Self-diagnosis was accepted due to the work of Rutter et al. \cite{rutter2023haven} who found self-diagnosis as a reasonable indicator of an internalised condition and to fall in line with similar works \cite{kasatskii2023effect}.

\subsection{Data Gathering}
The data collection took place between June and July 2024 following ethics approval from the Faculty of Science and Technology Ethics Committee at Lancaster University in May. Participants were given detailed instructions on what would take place from a script used by the researcher. Participants were given chance to ask any questions they might have before beginning the session. They then completed a coding task involving debugging and generating original code while the screen activity was being recorded. Once the coding activity was complete, the participant would watch the recording while narrating their thought process. They would then take a short break before filling out a Likert scale survey and then taking part in a short interview. The interviews discussed the coding activity and the students' interaction with the IDE. The session was recorded and automatically transcribed.

\subsection{Data Analysis}
Thematic analysis was used as described by Braun \& Clarke \cite{braun2023toward}. The transcripts were reviewed for accuracy of the auto transcription and prepared for analysis by marking when participants or the researcher were speaking. Multiple read-throughs were then conducted of the full corpus of nine transcripts, which contained 37965 words in total. Notes were made on observations of potential codes throughout until saturation was reached. The notes made were then reviewed and compiled into a list of open codes and axial codes to form a code book. Two researchers then independently coded a sample transcript. The transcript was selected at random from any of the transcripts which constituted at least 10\% of the total corpus using the code book \cite{oconnor20}. Once the sample was coded the two coders met and reviewed the coded transcripts to reach negotiated agreement on the code book before coding the remaining transcripts.

\begin{figure}[htp]
    \centering
    \includegraphics[width=1\linewidth, height=55mm]{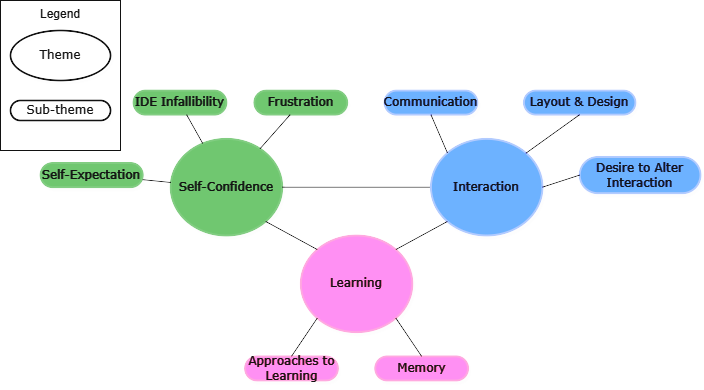}
    \caption{Diagram representing themes identified through thematic analysis}
    \label{fig:enter-label}
    \label{themes}
\end{figure}

Following the coding phase, we identified a number of themes as shown in Figure \ref{themes} in line with the methods of Braun and Clark \cite{braun2023toward}.

\section{Results}

\subsection{Main Themes Descriptions}
The theme of self-confidence demonstrates the difficulties students experienced when completing the coding task as part of the study and how their perceptions impacted their ability to complete it. 

The interaction theme highlighted aspects of the IDE which impacted students' ability to complete the task. 

The learning theme demonstrated the ways students engaged with learning in relation to the IDE and also barriers they faced to fully understanding and utilising the tool. 

\subsection{Self-confidence Sub-Themes}
\subsubsection{Self Expectation}
One of the major factors that was identified was the high level of self expectation participants placed on themselves. Participants in the study often seemed stressed or disheartened and harsh on themselves when they failed to meet their high standards.
\begin{quote}
    \textit{“And now I look like an idiot.”} (P04 Interview, describing how long it took them to get the code to run).
\end{quote}

The balance of self expectation and difficulty experienced was also visible when participants provided justifications for not meeting their desired standard. This need to clarify why they didn't perform as they would have liked was often self focused rather than allowing consideration for other factors such as usability.
\begin{quote}
    \textit{“But I basically just expected better of myself really. I should know this stuff. If I'd been doing a bunch more Java development in the last week or so, I probably would've been better at this, but I think that's just me being a bit rusty and feeling it should be a bit better than that.}
    
    \textit{…}
    
    \textit{If this was python we would have been gone 20 minutes ago”} (P08 Interview, describing their disappointment in themselves for not being able to complete the task quicker).
\end{quote}
This sub-theme also lead to participants focusing the blame on themselves for their mistakes rather than the IDE, even in situations where the IDE had been unclear or had not functioned as expected.
\begin{quote}
    \textit{“The coding environment was easy enough, myself is the problem.”} (P07 Interview, describing their belief that they were the reason they struggled in the task).
\end{quote}
\subsubsection{IDE Infallibility}
An assumption present throughout this study was students assuming the IDE must be working correctly and the fault was on them for how they were using the tool. The IDE infallibility could also be clearly seen in the Likert scale responses from Figure \ref{likertgraph}, with participants giving the IDE positive scores, despite qualitative statements in both the think aloud and interview contradicting those scores, such as Participant 3 who gave the design and layout of the IDE a high Likert score despite struggling with the layout.
\begin{quote}

    \textit{"So sometimes if I'm just trying to use a terminal and then I see on the side there's 10 terminals or something or there's just so many buttons that you can press. But I'm just trying to do something simple.”} (P03 Interview, describing being overwhelmed by the UI of the IDE).
\end{quote}
This theme was consistent and present throughout the study, with participants taking longer to solve issues due to being focused on their own work rather than potential issues with the IDE set up.

\begin{quote}
    \textit{“Yeah, I think I was more inclined to think it was probably me.”} (P03 Interview, describing how they assumed they were doing something wrong rather than the IDE not being set up correctly).
\end{quote}

\subsubsection{Frustration}
Frustration was evident throughout most of the activities, with participants being visibly frustrated and annoyed with both their own performance and the IDE failing to behave as expected. It was notable that participants frustrations would often grow rapidly and make it harder for them to focus, especially when not meeting their own standards.
\begin{quote}
    \textit{“It almost feels frustrating that it took you so long to figure something out that was such a simple fix and I was overthinking it in a way and I was ignoring certain details.”} (P01 Think aloud activity, describing how installing an extension fixed all the issues they had been having getting the IDE to run the code).
\end{quote}

\begin{quote}
    \textit{“And I don't think it's an error with my coding anymore and I'm trying to find the solution online but not being able to quickly locate the correct information to change what I'm doing and it becomes very frustrating quite quickly.”} (P07 Interview, describing trying to look up information on how to run the code in the IDE).
\end{quote}
Even participants who found the task simple recounted how outside of the study there were occasions where the IDE would cause frustration, despite the participant considering themselves skilled at using the tool.
\begin{quote}
    \textit{“Sometimes that can be more annoying if it's only got part of the validation rules right. Or if it's not set up properly.”} (P09 Interview, describing the experience of using an IDE when it is not working as expected).
\end{quote}

\subsection{Interaction Sub-Themes}
\subsubsection{Communication}
The interaction between the participants and the IDE was often limited and restricted by how well the IDE communicated with the participant. Its ability to communicate if it was running or not, or if it was set up correctly to compile the code was limited and often caused confusion for the participant.

\begin{quote}
    \textit{“And then yeah, I tried to start running and then nothing happened.”} (P03 Think aloud activity, describing trying to use the IDE to run the code).
\end{quote}

\begin{quote}
    \textit{“It just downloads a random folder for you and then I wasn't sure if you are meant to open the folder or if you are meant to move that folder elsewhere.}

    (P02 Interview, describing how the IDE downloaded a file to run the code without information on using the file).
    
\end{quote}

The communication of errors also caused confusion for participants, with it often being unclear how serious the warning was.
\begin{quote}
    \textit{“Yeah, I think I got stuck on that warning for a while. Although I think it's a bit confusing when it has that light bulb next to it. I know that it does the light bulb when it's something that Visual Studio could help resolve, whereas the other warnings are things that it can't find an easy fix for, but I think it was slightly confusing if that's an error or a warning.” }(P01 Think aloud activity, describing the similarity between warnings and other indications in the IDE).
\end{quote}

\subsubsection{Layout and Design}
It was frequently observed that the design and layout of the IDE had an impact on the experience of participants. This varied between participants with some finding the amount on the screen overwhelming, while some felt the IDE was not displaying enough information. The first aspect of design was how the pop ups and amount of activity on screen created a frequent susceptibility to overwhelm.
\begin{quote}
    \textit{“It's like you open it up and a bunch of stuff just pops up on the screen and it's really a little bit overwhelming.”} (P05 Think aloud activity, describing the how the IDE appears when first turned on).
\end{quote}
It was also highlighted that this sense of overwhelm discouraged participants from making use of helpful and sometimes essential tools for completing the task.
\begin{quote}
    \textit{“I get overwhelmed by a lot of information quite easily and I feel like the extensions, to me, feel very overwhelming and over stimulating. So I don't go to them so it never crossed my mind that the problem was with the extension rather than anything else.”} (P01 Think aloud activity, describing how they avoided the extensions needed to run the code due to finding them overwhelming).
\end{quote}

The difficulty of managing overwhelm while trying to find tools could be amplified by the difficulty of knowing what to look for. Participants described the interface as busy and cluttered, which made it difficult to notice when the IDE was trying to give useful information.
\begin{quote}
    \textit{“And a lot of tools that are available while they're supposed to be very useful aren't necessarily very clear, and it can become very easy to click on a button and think why has nothing happened and you've missed the little pop up window at the top that does everything because so many functions in Visual Studio lead to that pop up window where you then not only may miss it, but you also may already need to know what you need to type to get the next thing to happen.”} (P07 Interview, describing the issues with keeping track of important pop ups in the IDE).
\end{quote}

\subsubsection{Desire to Alter the Interaction}
This sub-theme highlights the consistent desire to alter the design or layout of the IDE. This could be seen with the participants wanting to change what was on screen at different times and again there was not one standard desire with variations of participants wanting more or less on screen or only specific things. This sub-theme was most prevalent in how participants wanted information to be shown, such as the position on the screen or being able to see multiple aspects of code at once.
\begin{quote}
    \textit{“I split the my code into the other bit so I could process it.”} (P06 Think aloud activity, describing using the split screen function in the IDE to look at different code files at the same time).
\end{quote}

There was also a prevalent feeling of wanting to change or reduce what information was visible and to have it limited to things useful at the time.
\begin{quote}
    \textit{“Because if you minimise everything and if you neaten it just because it looks nicer. But if you minimise everything then you can go to the important classes and if you do your find the methods it'll take you straight to the methods.”} (P02 Interview, describing how they like to minimise the amount of things on screen at once while coding).
\end{quote}

\subsection{Learning Sub-Themes}
\subsubsection{Approaches to Learning}
The approaches displayed by participants highlighted the different ways users might try and solve problems they could be faced with in an IDE. The sub-theme of approaches to learning was put prominently upfront by the study. This prominence was partially due to the frequent feeling that there was a knowledge barrier to using the tool, which could not be overcome by the guidance the tool provided. These barriers often required peer support or independent research when other help was unavailable.
\begin{quote}
    \textit{“At first I found there was a lot that you had to do to get it ready to go and if you didn't have someone there to explain some of it to you, you do have to Google it to discover the information.” }(P07 Interview, describing the difficulty of getting the IDE set up).
\end{quote}
When conducting this research participants had to overcome other aspects of difficulty, such as navigating information which was outdated due to updates and changes to the IDE tool, and would sometimes seek visual confirmation to help in their research to identify relevant information.
\begin{quote}
    \textit{“So something that you find online that's talking about a certain UI might be different to the UI that you have currently. I usually try and find images to make sure that there are similarities.”} (P01 Think aloud activity, describing trying to find solutions to get the IDE working while navigating solutions online which are different versions of the IDE).
\end{quote}
Participants would struggle with navigating dense sources of information which were difficult to parse or would be challenging for the participant to quickly review for useful information. This difficulty in navigating the information would result in the participant missing useful information.
\begin{quote}
    \textit{“Albeit when I do my research I kind of skip over a lot of stuff that may be helpful to me.”} (P01 Think aloud activity, describing how they tend to skip over big pieces of text trying to find a quick solution).
\end{quote}

\subsubsection{Memory}
This sub-theme is perhaps most relevant to the area of learning for people with ADHD as they often struggle with memory \cite{BarkleyRussellA2018Ahd}. The sub-theme highlights the difficulty of balancing learning multiple coding languages and remembering how to navigate a complex interface. One aspect of memory that appeared as difficult for participants was balancing negative interaction experiences with the tool and keeping track of the different variables and functions as part of the coding task.
\begin{quote}
    \textit{“I think with ones that don't suit my style of work and don't conform to how I read and interact with things, it can be very difficult. And because I have to bounce around and re-remember where everything is every couple of seconds.” }(P05 Interview, describing how different interfaces can make it more difficult to keep track of things when coding).
\end{quote}

The most prevalent impact on memory was remembering specifics of Java when they potentially hadn’t used it recently, which is common when learning how to programme and completing different courseworks in different coding languages.

\begin{quote}
    \textit{“And I'm used to that whereas well mainly because I haven't used Java in a while so that happened.”} (P04 Interview, describing how they are used to the run button being different in different IDEs and being out of practice with the way to run Java code).
\end{quote}
The above also amplified some of the earlier aspects of self-confidence that have been discussed with struggling to remember leading to participants blaming themselves for not being able to get the tool to work.
\begin{quote}
    \textit{“I think I just assumed I didn't know how to run it. I'd just forgotten everything with coding.”} (P02 Interview, describing their confusion over the code not running before the IDE was set up correctly but assuming they had done something wrong).
\end{quote}

\section{Discussion}

\subsection{Self-confidence}
The presence of self-confidence as a major theme of a study involving ADHD is not surprising. Existing literature discusses how people with ADHD often have issues with self-esteem \cite{newark2016self}. Bodalski et al. \cite{bodalski2023adhd} also discuss the relationship between self-esteem and procrastination in students with ADHD. Self-confidence was found to play a considerable role in how students interacted with the IDE and their experience while using it. Issues such as self-deprecation, self-doubt and worrying about or losing track of time were some of the prominent observations. The presence of the observations in line with the established understanding from literature highlights the weight that should be given to supporting students with ADHD and their difficulties with self-confidence.
However, insights from this study show that while the theme of self-confidence was present, the design of the IDE seemed to amplify this phenomenon.
\begin{quote}
    \textit{"The IDE wasn't complete or set up properly and I'd assumed that there was something wrong in the code that I had or I'd missed another bug or something."} (P05 Interview, describing assumptions that something must still be wrong with the code when the IDE wasn't running it).
\end{quote}
Instances of the IDE not being clear on its state of readiness lead to students being untrusting of the code they had generated rather than considering other factors at play. The implications of the impacts from the current design leads to the potential aggravation of difficulties already faced by students with ADHD. 
The findings around the interaction of self-confidence and the current design of IDEs would indicate the need for further research in how we can design tools to support this prevalent factor.

\subsection{Interaction}
The nature of interaction was a prevalent theme throughout the study. The various preferences and feelings regarding aspects of the design and use of the IDE appeared in both the think aloud exercise and the interview. A common occurrence was participants feeling there was too much happening on screen, a phenomenon also observed more generally by Seery et al. \cite{seery2025one} for users with ADHD. Understanding the accessibility impact of IDE design is especially important due to the negative impact that can be had on disabled people's ability to engage when technology is not equally usable for them \cite{shaheen2019bringing}.

Adaptability and being able to personalise the interaction were major factors. This falls in line with the work of Kasatskii et al. \cite{kasatskii2023effect} who discuss the importance of control over perceptual load for individual software engineers. Making it easier for students to influence their interaction with the IDE could improve their overall learning experience.

\subsection{Learning}
The theme of learning was evident through students’ varied approaches to solving a coding problem, including how and to what extent prior experience was utilised. This theme interacted heavily with the themes of self-confidence and interaction, which has been documented in other literature. Di Lonardo Burr \& LeFevre for example, discuss the impact of self-confidence on learning \cite{DILONARDOBURR2020101808} and highlight the importance of trying to mitigate detrimental factors to self-confidence by creating a positive learning environment.

How participants approach the learning aspect of problem solving was closely linked to how they used the IDE. Some participants were heavily dependent on search tools or trial and error, while others focused on making notes in the comments and using those to "rubber duck" the problem. Some participants also discussed how they might find alterations to the IDE or extra features beneficial towards their approaches to learning.

\begin{quote}
    \textit{"If I can put notes on like oh this function does this thing, you should use this one or this is the function you're looking for if I'm hovering over something that might be quite useful."} (P08 Interview, describing how they might find it useful to be able to have a note maker built into the IDE to keep track of different information).
\end{quote}

In the previous theme of interaction, adaptability was identified as a factor which could support accessibility. The learning theme endorses this concept, displaying adaptability as an avenue to better integrate the IDE in students’ approaches to learning. The desire of participants to personalise the IDE to their approaches to learning highlights a key consideration for making IDEs digitally accessible, and a positive learning environment.

\subsection{Limitations and opportunities}

The primary limitation of this work would be that the sample size for the study conducted is relatively small. While the sample size is somewhat standard for a study of this nature, it still leaves opportunities for future work to build on these findings with larger groups of participants.

\section{Conclusion}
We set out the following research questions which drove our research:

\textbf{RQ1.}	Does  the UI design and tools within  an IDE have an impact on ease of use for a student with ADHD?

\textbf{RQ2.}	What are the experiences  of students with ADHD when using an IDE?

Our findings show that the answer to RQ1 would be that yes, the design and tools within an IDE have an impact on students with ADHD. The impact being that layout and usability can be a barrier for the students and impact their ability to complete tasks and build confidence in coding. 

Further, we identified various experiences when using an IDE but primarily identified frustration and that students had difficulty making full use of the tool leading to an overall negative experience for the students.

Finally the identified themes provide insight into the user experience of computing students with ADHD and helpful information for those who design digital tools for them. In future work we anticipate building on these findings to guide a design process addressing and understanding the best way to support students with ADHD in their coding education. We believe by using these themes in further participatory research, researchers will be able to understand how to create accessible and usable IDEs for students with ADHD in a learning setting.


\section{Acknowledgements}
Thanks to Professor Joe Finney for providing the initial code base for the coding task and George Limbert for assistance in the open coding process.

\bibliographystyle{splncs04}
\bibliography{mybibliography}

\end{document}